\documentclass{svjour2}                    
\usepackage{hyperref}

\smartqed  
\usepackage{graphicx,comment}
\usepackage{color,ulem,amssymb,amsmath}
\definecolor{greencolor}{rgb}{0,0.5,0.2}
\definecolor{redcolor}{rgb}{.7,0.,0.}
\definecolor{bluecolor}{rgb}{0,0.,1.}
\definecolor{greycolor}{rgb}{.5,.5,.5}

\begin{document}

\title{Quantitative universality for a class of weakly chaotic systems}

\author{Roberto~Venegeroles}

\institute{Roberto Venegeroles \at
              Centro de Matem\'atica, Computa\c c\~ao e Cogni\c c\~ao\\
 Universidade Federal do ABC, 09210-170, Santo Andr\'e, SP, Brazil
\\
              \email{roberto.venegeroles@ufabc.edu.br}          
}

\date{Received: date / Accepted: date}

\maketitle

\vspace{-0.5cm}
\begin{abstract} 
We consider a general class of intermittent maps designed to be weakly chaotic, i.e., for which the separation of trajectories of nearby initial conditions is weaker than exponential. We show that all its spatio and temporal properties, hitherto regarded independently in the literature, can be represented by a single characteristic function $\phi$. A universal criterion for the choice of $\phi$ is obtained within the Feigenbaum's renormalization-group approach. We find a general expression for the dispersion rate $\zeta(t)$ of initially nearby trajectories and we show that the instability scenario for weakly chaotic systems is more general than that originally proposed by Gaspard and Wang [Proc. Natl. Acad. Sci. USA {\bf 85}, 4591 (1988)]. We also consider a spatially extended version of such class of maps, which leads to anomalous diffusion, and we show that the mean squared displacement satisfies $\sigma^{2}(t)\sim\zeta(t)$. To illustrate our results, some examples are discussed in detail.
\end{abstract}

\section{Introduction}

The instability of deterministic motion is typically characterized by the growing separation of initially nearby trajectories, which, in turn, indicates that dynamics is high sensitive to its initial state; very small differences at one moment in such systems can result in very large differences later on. For a chaotic dynamical system $x_{t}=T^{t}(x_{0})$, the separation of initially nearby trajectories, namely $\delta x_{t}=T^{t}(x_{0}+\delta x_{0})-T^{t}(x_{0})$, evolves asymptotically as
\begin{equation}
\label{wc}
|\delta x_{t}|\sim|\delta x_{0}|\exp[\Lambda_{t}(x_{0})\zeta(t)],
\end{equation}
for almost every choice of $\delta x_{0}$ within any infinitesimal neighborhood of $x_{0}$. The positive coefficient $\Lambda_{\infty}(x_{0})$ stands for the largest Lyapunov exponent for $\zeta(t)\sim t$, and most of well-known chaotic systems are ruled by dispersion rates $\zeta(t)$ of this type \cite{Ose,BGGS,WSSV}. After the pioneering work of Gaspard and Wang \cite{GW}, there has been in recent years a growing interest in understanding weakly chaotic (sporadic) systems for which the separation of initially nearby trajectories is weaker than exponential, i.e. $\zeta(t)$ sublinear \cite{BIG,RZ,KB,AA,PSV,SV1,RV,RK,SVpr}. For these systems, the conventional Lyapunov exponent vanishes and the weakly chaotic behavior results from the intermittent switching between long regular phases (so-called laminar) and short irregular bursts.

We consider here a general class of maps of the interval which are weakly chaotic according to
\begin{equation}
\label{wcc}
\zeta(t)\sim l(t)t^{\gamma},\qquad 0\leq\gamma\leq 1,
\end{equation}
being $l(t)$ is a slowly varying function at infinity so that $l(t\rightarrow\infty)=\infty$ for $\gamma=0$ and $l(t\rightarrow\infty)=0$ for $\gamma=1$. Then, we tie all of its spatio and temporal properties, including the map equations itself, to a single characteristic function $\phi$. The main step towards a unifying framework is based on determining the eigenfunctions of Feigenbaum's renormalization operator \cite{Feig1,Feig2}
\begin{equation}
\label{feig}
\mathcal{F}g(x)=\alpha g[g(x/\alpha)],
\end{equation}
where $\alpha$ is a rescaling factor. The eigenfunction that defines the universality class is a fixed point of $\mathcal{F}$, i.e. $\mathcal{F}g_{*}=g_{*}$.  We shall see that
\begin{equation}
\label{phiuniv}
g_{*}(x)=\phi^{-1}[\phi(x)-\tau],
\end{equation}
where $\phi^{-1}$ denotes the inverse of $\phi$ and $\tau$ is a constant.

Since the early 1980's, the idea that the same functional equation employed by Feigenbaum for studying the period-doubling cascade can also be used to describe intermittency and dissipative systems has been investigated by many authors, notedly \cite{ETW,HNS,HR,PS,Kuz,FKS}. From a viewpoint more connected to the intermittency phenomenon, considerations so far about universality have been based on the scaling properties of the laminar length \cite{ETW,HNS,HR,PS}. Our results also shed some light on the scaling hypothesis behind the relationship between laminar length and Feigenbaum's operator (\ref{feig}).

We provide a general formula for the dispersion rate $\zeta(t)$, one of the main results of this manuscript. But besides an attempt to outline the subexponencial instability in weak chaos, our goal is to provide a full description of a nonlinear dynamical system in which a scaling property is present. Here we have two key quantities determining the spatio and temporal properties of weakly chaotic systems. The invariant density $\rho(x)$ gives us the measure of concentration trajectories at each stage of intermittency, whereas the residence times at each stage are ruled by the waiting-time probability density function $\psi(t)$ of the laminar region. By introducing a proper modeling of intermittency mechanism, together with a renormalization-group approach, we establish a universal criterion for the choice of $\phi$ which enables us to predict the dispersion rate $\zeta(t)$, as well as determining $\rho(x)$ and $\psi(t)$. We will also see that these results enable us to predict the anomalous subdiffusion for spatially extended versions of such systems by means of a relationship between $\zeta(t)$ and the mean squared displacement $\sigma^{2}(t)$.

Subexponential instability (\ref{wcc}) is a consequence of the infinite invariant measure $\int d\mu(x)=\int\rho(x)dx=\infty$ over the laminar region, and we will establish here this relationship quantitatively. On the other hand, the stagnant motion of laminar region is also related to Eq. (\ref{wcc}), leading to an algebraic behavior of waiting-time probability density as we shall see later on. In Hamiltonian systems, such behavior corresponds to the tendency of nonescaping particles to concentrate around regular regions, such as stable islands \cite{MMGK,RVc,RVu,AN} and invariant tori of nearly-integrable systems \cite{Aiz,MV}. In particular, a $\gamma=0$ weakly chaotic model has been recently used to model the Nekhoroshev stability \cite{Nek}, see \cite{SA}.

\section{Model and Modeling}

Let us introduce the general class of piecewise $C^{2}$ expanding maps, from $[0,1]$ to itself, in the form $x_{t+1}=T(x_{t})$ so that
\begin{equation}
\label{map}
 T(x)=x+f(x)
\end{equation}
on $(0,c)$, with a single marginal fixed point at $x=0$, i.e.
\begin{equation}
\label{indff}
 f(x\rightarrow 0)=f'(x\rightarrow 0)=0
\end{equation}
and $T(c_{-})=1$. Although the global form of $T$ is less relevant here, the mechanism of intermittency also relies on the existence of a second branch such that $T:(c,1)\rightarrow(0,1)$, ensuring the chaotic reinjection of trajectories to the laminar region near the marginal point. The best known model that meets these criteria are Pomeau-Manneville type maps $T(x)=x+ax^{1+1/\gamma}$ mod 1, where $a$ and $\gamma$ are positive parameters \cite{Mann,PM}. This and others examples will be discussed in the next sections.

One key ingredient for the resulting weakly chaotic behavior (\ref{wcc}) is how invariant density of $T$ behaves near the marginal fixed point. First of all, $\rho(x)$ must be an eigenfunction of the Perron-Frobenius operator (see for instance \cite{HS})
\begin{equation}
\label{PF}
\rho(x)=\sum_{j}\frac{\rho[T_{j}^{-1}(x)]}{|T'[T_{j}^{-1}(x)]|},
\end{equation}
where the sum extends over all preimages $T_{j}^{-1}$ of the point $x$ at which the density is to be computed. Near the marginal fixed point $x=0$ we have $T_{0}^{-1}(x)\sim x$, whereas $T_{j\neq0}^{-1}(x)$ assumes values on the interval far away from zero. Then one has $\rho(x)\sim\rho(x)[1-f'(x)]+\mbox{constant}$, resulting in the divergent invariant density
\begin{equation}
\label{deninv}
\rho(x)\sim\frac{1}{f'(x)}, 
\end{equation}
up to a positive multiplicative constant for non-normalizable $\rho(x)$.

The finite-time (generalized) Lyapunov exponent of the map $T$ satisfying Eq. (\ref{wc}) is
\begin{equation}
\label{laver}
\Lambda_{t}(x_{0})=\frac{1}{\zeta(t)}\sum_{k=0}^{t-1}\ln|T'[T^{k}(x_{0})]|.
\end{equation}
In order to obtain the dispersion rate $\zeta(t)$, we expect that the average of Lyapunov exponent (\ref{laver}) over the initial condition ensemble may be estimated by using a density function $\rho(x,t)$ so that 
\begin{equation}
\label{laver1}
\left\langle \Lambda\right\rangle=\lim_{t\rightarrow\infty}\frac{1}{\zeta(t)}\int_{0}^{t}du\int_{0}^{1}dx\,\rho(x,u)\ln|T'(x)|.
\end{equation}
To find $\rho(x,t)$ we will consider the continuous-time stochastic model proposed in \cite{IGR},
\begin{equation}
\label{pde}
\frac{\partial}{\partial t}\rho(x,t)=-\frac{\partial}{\partial x}[v(x)\rho(x,t)]+C(t).
\end{equation}
The convective equation (\ref{pde}) describes the stochastic motion of a particle initially in the laminar phase until it is random reinjected back to a position on the interval after reaching (crossing) the point $x=c$. The term $v(x)$ describes the particle's velocity in the laminar phase, and the reinjection source term $C(t)$ is chosen to fulfill conservation of measure $\int_{0}^{1} dx\,\rho(x,t)$. Equation (\ref{pde}) can be completely solved using the method of characteristics \cite{MC}. Assuming uniform initial density $\rho(x,0)=1$ and considering the characteristic function $\phi(x)$ as follows
\begin{equation}
\label{phi}
-\int\frac{1}{v(x)}dx=\phi(x),
\end{equation}
we get the general solution for the Laplace transform $\mathcal{L}_{s}[\rho(x,t)]=\tilde{\rho}(x,s)$ (see Appendix A1):
\begin{equation}
\label{solve}
\tilde{\rho}(x,s)=\frac{1}{v(x)}\frac{\tilde{\psi}[s+\phi(x)]}{1-\tilde{\psi}[s+\phi(1)]},
\end{equation}
where the corresponding waiting-time density function $\psi(t)$ is (hereafter omitting positive multiplicative constants)
\begin{equation}
\label{psi}
\psi(t)=-[\phi^{-1}(t)]'.
\end{equation}

Equation (\ref{solve}) is consistent with conservation of $\int_{0}^{1} dx\,\rho(x,t)$, see Appendix A2 for details. Note that $\phi$ in Eq. (\ref{phi}) is defined up to an integration constant, which is chosen so that $\tilde{\psi}(\phi(1))=1$ (normalization), and thus $\phi(1)=0$. The average of (generalized) Lyapunov exponent is so that $\langle\Lambda\rangle\propto\int_{0}^{1}dx\rho(x)\ln|T'(x)|$, and therefore we must have the scaling $\rho(x,t)\sim\zeta'(t)\rho(x)$ as $t\rightarrow\infty$, see Eq. (\ref{laver1}) and, for more details, see also Appendix A3. Upon considering this asymptotic behavior, we must have
\begin{equation}
\label{vxrho}
v(x)\sim\frac{x}{\rho(x)}.
\end{equation}

The dispersion rate $\zeta(t)$ can now be calculated from the general solution (\ref{solve}), and one gets
\begin{equation}
\label{sepwe}
\zeta(t)\sim\mathcal{L}_{t}^{-1}\left\{\frac{1}{s[1-\tilde{\psi}(s)]}\right\}.
\end{equation}
Later we will solve Eq. (\ref{sepwe}) explicitly in terms of $\phi$. Besides the waiting-time density (\ref{psi}), the map $T$ and its corresponding invariant density can also be put forward here in terms of $\phi$ as follow
\begin{equation} 
\label{nn}
T'(x)\sim 1-\frac{1}{x\phi'(x)},\qquad \rho(x)\sim-x\phi'(x).
\end{equation}

\section{Infinite Invariant Measure Implies Subexponential Instability}

Now we are ready to carry out the renormalization-group approach by using the length of laminar motion $x_{*}$. A detailed justification for this choice will be given here later on. From the definition of $\phi$ in Eq. (\ref{phi}) we have the relation $\tau=\phi(x)-\phi(x_{*})$, $\tau$ being the time interval between laminar positions $x_{*}$ and $x$. Thus the eigenfunction of renormalization operator (\ref{feig}) is $g_{*}(x)=x_{*}$, as presented before in Eq (\ref{phiuniv}). The boundary conditions must be such that $\phi'(x\rightarrow0)=-\infty$ since we must have $T'(x\rightarrow0)=1$ and positive invariant density. Thus $g_{*}(x)$ has the same characteristics of map $T$ for $x$ near zero:
\begin{equation}
\label{bdg}
g_{*}(x\rightarrow 0)=0,\qquad g'_{*}(x\rightarrow 0)=1.
\end{equation}

For our first example, first notice that the eigenfunction $g_{*}(x)$ of operator (\ref{feig}) can be recast in the form $g_{*}(x)=\alpha\phi^{-1}[\phi(x/\alpha)-2\tau]$. Introducing the auxiliary functions $h_{1}(x)=\alpha\phi^{-1}(2x)$ and $h_{2}(x)=\phi(x/\alpha)/2$, one has $g_{*}(x)=h_{1}[h_{2}(x)-\tau]$, admitting as possible solutions $h_{1}=\phi^{-1}$ and $h_{2}=\phi$. Therefore, the characteristic function $\phi$ satisfies the relations $\phi(x/\alpha)=2\phi(x)$ and $\alpha\phi^{-1}(2x)=\phi^{-1}(x)$. The first of latter two equations was derived in a different way in \cite{HR}, and it is easy to see that the pair admits as solution
\begin{equation}
\label{ph2}
\phi(x)=x^{-1/\gamma},\qquad \alpha=2^{\gamma},
\end{equation}
for $\gamma>0$. This result gives us the well-known class of Pomeau-Manneville maps, for which $f(x)\sim x^{1+1/\gamma}$ near $x=0$. From the physical point of view, the original Pomeau-Manneville system, i.e. $\gamma=1$, is paradigmatic since it corresponds to certain Poincar\'e sections related to the Lorenz attractor \cite{PM}. Lastly, one has $\rho(x)\sim x^{-1/\gamma}$ and  $\psi(t)\sim t^{-(1+\gamma)}$. Our Eq. (\ref{sepwe}) gives us the corresponding weakly chaotic regimes:
\begin{equation}
\label{bern}
\zeta(t) \sim \left\{
\begin{array}{ll}
 t^{\gamma},  & \,0 < \gamma < \displaystyle 1,\\
t/\ln t, & \displaystyle \qquad\gamma = 1,
\end{array}
\right.
\end{equation}
and also $\zeta(t)\sim t$ for $\gamma>1$.

The weakly chaotic instability (\ref{wcc}) stems from the diverging behavior of invariant measure $\mu$ near the marginal fixed point, e.g. $0<\gamma\leq1$ for the class of Pomeau-Manneville maps (\ref{ph2}). From Eq. (\ref{nn}) we have the invariant measure $\mu([x,y])\sim x\phi|_{y}^{x}-\int_{y}^{x}du\phi(u)$. Therefore, the divergence of invariant measure occurs provided that the characteristic function $\phi$ obeys
\begin{equation}
\label{cond}
\int dx\phi(x)\rightarrow-\infty,\qquad x\rightarrow0,
\end{equation}
which is more restrictive than the boundary conditions (\ref{bdg}). From Eq. (\ref{psi}) we also have the relation $\int dt\psi(t)t\sim-\int dx\phi(x)$, and therefore Eq. (\ref{cond}) implies the divergence of mean waiting-time. Thus, the weakly chaotic behavior occurs provided that $\psi(t)$ does not decrease faster than $t^{-2}$. In such cases, Eq. (\ref{sepwe}) can be solved by making use of Karamata's Abelian and Tauberian theorems for the Laplace-Stieltjes transform \cite{Feller}. By considering the general form of cumulative distribution function associated to $\psi(t)$, i.e., $\int_{0}^{t}du\psi(u)\sim1-1/q(t)t^{\gamma}$, $q(t)$ being a slowly varying function at infinity, one obtains from Eq. (\ref{sepwe})
\begin{equation}
\label{kab}
\zeta(t)\sim\int_{0}^{t}du\mathcal{L}_{u}^{-1}\left[s\frac{q(1/s)}{s^{1+\gamma}}\right]\sim q(t)t^{\gamma},\qquad 0\leq\gamma<1.
\end{equation}
For $\gamma=1$ we have $\tilde{\psi}(s)\sim1-s|\ln s|/|q(1/s)|$, and thus Eq. (\ref{sepwe}) gives us $\zeta(t)\sim tq(t)/\ln t$. Furthermore, noting that $\int_{0}^{t}du\psi(u)\sim1-\phi^{-1}(t)$, one finally gets
\begin{eqnarray}
\label{zetaf}
\zeta(t)\sim \frac{1}{\phi^{-1}(t)}\times\left\{
\begin{array}{ll}
  1,  & \,0\leq\gamma < \displaystyle 1, \\
  1/\ln t, &\qquad \displaystyle \gamma=1.
\end{array}
\right.
\end{eqnarray}
Notice that Eq. (\ref{zetaf}) yields the same result (\ref{bern}) for Pomeau-Manneville's characteristic function (\ref{ph2}).

\section{Applications}

Our result (\ref{zetaf}) enables us to develop models with weakly chaotic behavior provided that the criterion (\ref{cond}) is fulfilled. Thus, our second example relies on the family of maps for which $\phi(x)$ behaves as
\begin{equation}
\label{phimap}
\phi(x)\sim\phi_{0}(x)\exp(x^{-\beta})
\end{equation}
for $\beta>0$, provided $\phi_{0}(x)$ does not go to zero as fast as $\exp(-x^{-\beta})$ for $x\rightarrow0$. This model should be understood as a $\gamma\rightarrow0$ limiting case for the rescaling factor $\alpha=2^{\gamma}$. Using again the machinery we have developed, one has $\psi(t)\sim t^{-1}(\ln t)^{-(1+1/\beta)}$, irrespective of $\phi_{0}$, resulting in the strong anomaly dispersion rate
\begin{equation}
\label{zetsl}
\zeta(t)\sim\ln^{1/\beta}t,
\end{equation}
which, together with Eq. (\ref{bern}), also agrees with formula (\ref{zetaf}).

The dispersion rates (\ref{bern}) and (\ref{zetsl}) are in perfect agreement with their corresponding quantities in the infinite ergodic theory \cite{ZT,SA1}, the so-called return sequences $a_{t}$ \cite{Aaronson}. Such sequences ensure a suitable time-weighted average of observables that converge in distribution terms towards a Mittag-Leffler distribution \cite{Aaronson}. Note also that our Eq. (\ref{deninv}) generalizes the invariant densities obtained in \cite{Th1,Th2} for these types of systems. Lastly we observe that, under condition (\ref{cond}), the dispersion rate (\ref{zetaf}) shows that the instability scenario for weakly chaotic systems is more general than that originally proposed by Gaspard and Wang in \cite{GW}. In particular, we can also consider weakly chaotic models with dispersion rates that grow faster than logarithms but slower than polynomials such as
\begin{equation}
\label{lnot}
\zeta_{a,b}(t)\sim\exp[b(\ln t)^{a}(\ln\ln t)^{1-a}]
\end{equation}
for $0\leq a<1$ and $b>0$ or $a=1$ and $b=\gamma$. Results (\ref{bern}) and (\ref{zetsl}) are respectively given by $\zeta(t)\sim\zeta_{1,\gamma}(t)$ for $0<\gamma<1$ and $\zeta(t)\sim\zeta_{0,1/\beta}(t)$. According to Eq. (\ref{zetaf}), the $0<a<1$ intermediary cases  give us
\begin{eqnarray}
\label{gsol}
\phi_{a,b}(x)&\sim&\exp\exp\left\{\frac{1-a}{a}\left[W\left(\frac{a}{1-a}\ln^{\frac{1}{1-a}}x^{-\frac{1}{b}}\right)\right]\right\}\nonumber\\
&=&O(\exp(\ln^{1/a}x^{-1/b})),
\end{eqnarray}
as $x\rightarrow0$, being $W$ the Lambert function \cite{WF}.

Yet another application of the renormalization-group approach: weakly chaotic maps of the type (\ref{map}) have been extensively used in the literature to model systems that exhibit anomalous transport, see for instance \cite{RK} and references therein. The mechanism for generating deterministic subdiffusion is based on the extended version of map (\ref{map}), from $[0,1/2)$ to the entire real line, according to the rules $f(x+N)=f(x)+N$ and $ f(-x)=-f(x)$, where $N$ assumes integer values. This results in a series of lattice cells with marginal points located at $x=N$. The corresponding transport properties can be understood in terms of a continuous-time random walk picture of this model, with probability density of waiting-times $\psi(t)$ near each marginal point. The mean squared displacement $\sigma^{2}(t)$ for such model is given by \cite{SM}
\begin{equation}
\mathcal{L}_{s}\{\sigma^{2}(t)\}\sim\frac{\tilde{\psi}(s)}{s[1-\tilde{\psi}(s)]}.
\end{equation}
Since $\tilde\psi(s\rightarrow0)=1$, from Eq. (\ref{sepwe}) one has
\begin{equation}
\label{sigz}
\sigma^{2}(t)\sim\zeta(t).
\end{equation}
The mean squared displacements for the extended versions of models (\ref{ph2}) and (\ref{phimap}) based on our results are in perfect agreement with those obtained respectively in \cite{GT} and \cite{HW,DK}. Equation (\ref{sigz}) is particularly interesting because it is a non-trivial extension for weakly chaotic systems of a relationship typically observed in usual chaos, namely $\sigma^{2}(t)\sim\zeta(t)\sim t$.

\section{Some Remarks on Feigenbaum's Scaling}

Why does the laminar length $x_{*}$ scale according to the Feigenbaum renormalization operator? The extension of the renormalization operator for the class of systems discussed here can be understood by means of the scaling limit
\begin{equation}
\label{lims}
g(x)=\lim_{n\rightarrow\infty}\alpha^{n}T^{2^{n}}(x/\alpha^{n}),
\end{equation}
since we have $T^{n}(x)\sim x+nf(x)$ near zero. By using recursion, it is simple to see that Eq. (\ref{lims}) leads to the renormalization operator (\ref{feig}) with $g_{*}(x)\sim x+\mbox{const.}f(x)$ and $\alpha=2^{\gamma}$. It is important to emphasize here that we need not find eigenfunctions of Eq. (\ref{feig}) covering the whole interval $[0,c]$, but just a vicinity of zero. Thus, we can expand our proposal eigenfunction (\ref{phiuniv}) for $\tau\approx0$ since $\phi(x)$ is singular at $x=0$, resulting $g_{*}(x)\sim x-\tau/\phi'(x)$. Now, from Eq. (\ref{nn}), one has $xf'\sim-1/\phi'$, and our scaling hypothesis boils down simply to $xf'\sim f$, in agreement with our hypothesis (\ref{indff}). Notice that although weakly chaotic systems (\ref{wcc}) are ruled by the same renormalization operator employed in the logistic map \cite{Feig1,Feig2}, what fundamentally distinguishes one case from the other are the boundary conditions behind their eigenfunctions.

A matter that deserves to be revisited in the literature is the perturbation analysis of the renormalization operator \cite{HNS,HR} since only the behavior of eigenfunctions near the marginal point is relevant here. Consider a perturbation of an eigenfuncion $g_{*}$, i.e., $g(x)=g_{*}(x)+\epsilon h_{\lambda}(x)$. The action of $\mathcal{F}$ on $g$ is so that $\mathcal{F}g(x)=g_{*}(x)+\lambda\epsilon h_{\lambda}(x)+O(\epsilon^{2})$, where $h_{\lambda}(x)$ satisfies the linearized renormalization operator $L_{\mathcal{F}}[h_{\lambda}(x)]=\lambda h_{\lambda}(\alpha x)$, given by
\begin{equation}
\label{pert1}
L_{\mathcal{F}}[h_{\lambda}(x)]=\alpha\{g'_{*}[g_{*}(x)]h_{\lambda}(x)+h_{\lambda}[g_{*}(x)]\}.
\end{equation}
After considering the boundary conditions (\ref{bdg}), Eq. (\ref{pert1}) boils down simply to $2\alpha h_{\lambda}(x)=\lambda h_{\lambda}(\alpha x)$ for $x\sim0$, which implies homogeneity of degree $p$ for $h_{\lambda}$ and, therefore, $\lambda=2^{1+\gamma(1-p)}$. Note also that successive applications of $\mathcal{F}$ on $g$ are such that
\begin{equation}
\label{pert2}
\mathcal{F}^{n}g(x)=g_{*}(x)+\sum_{\lambda}\lambda^{n}c_{\lambda}h_{\lambda}(x),
\end{equation}
and, thus, the stability condition imposes $p>1+1/\gamma$. This means the invariance of the class of maps (\ref{map}) under the symmetry $T_{\gamma}\mapsto T_{\gamma}+O(f_{\gamma})$ near $x=0$, which is perfectly consistent with the results developed here. Interestingly, the perturbation  (robustness) analysis does not distinguish weak chaos from usual chaos, i.e., there is no symmetry breaking at $\gamma=1$, despite there being a phase transition at this value (see \cite{RV} and references therein).

\section{Conclusions}

The invariant density $\rho(x)$, the waiting-time probability density function $\psi(t)$ and, even more importantly, the dispersion rate of initially nearby trajectories $\zeta(t)$, are all described here by a single characteristic function $\phi$. We show that this function is closely related to the Feigenbaum renormalization-group operator (\ref{feig}). By means of an inverse problem approach we show that, given a choice of $\phi$ satisfying Eq. (\ref{cond}), all of these fundamental quantities automatically become known, including the map equations itself. Thus all of these results, namely Eqs. (\ref{psi}), (\ref{nn}), (\ref{zetaf}), and (\ref{sigz}), together with Eq. (\ref{wcc}), unify a paradigmatic class of weakly chaotic systems, the most general hitherto known, in a simple and powerful way.

We believe that the main question raised in \cite{GW}, i.e. whether intermediate dynamical behaviors of the type (\ref{wcc}) could exist in the range $0\leq\gamma\leq1$, was reasonably elucidated here. In particular, we propose a broad class of weakly chaotic models with dispersion rates $\zeta(t)$ that grows faster than logarithms but slower than polynomials, also covering these bounding cases for appropriate choices of parameters, see Eqs. (\ref{lnot}) and (\ref{gsol}).

\begin{acknowledgements}
The author gratefully acknowledges the helpful discussions with Alberto Saa. This work was supported by the Brazilian agencies CNPq and FAPESP.
\end{acknowledgements}

\section*{Appendix}

\subsection*{\bf A1. General solution}

Equation (\ref{pde}) can be completely solved by using the method of characteristics \cite{MC} for the homogenous density $\rho_{h}(x,t)$, where $\rho(x,t)=\rho_{h}(x,t)+\rho_{n}(x,t)$. For uniform initial density $\rho_{h}(x,0)=1$ one has
\begin{equation}
\label{rhoh}
\rho_{h}(x,t)=\frac{1}{v(x)}\psi(t+\phi(x)),
\end{equation}
where
\begin{equation}
\label{psiv}
\psi(u)=v(\phi^{-1}(u)),
\end{equation}
together with Eq. (\ref{phi}). From Eqs. (\ref{phi}) and (\ref{psiv}) one has Eq. (\ref{psi}). The nonhomogeneous term is given by
\begin{equation}
\label{nhom}
\rho_{n}(x,t)=\int_{0}^{t}d\tau\,C(\tau)\rho_{h}(x,t-\tau).
\end{equation}
The source term $C(t)$ is so that $\int_{0}^{1}dx\rho(x,t)$ is conserved. After introducing this condition in Eq. (\ref{pde}) and also considering that we shall have Eq. (\ref{vxrho}), i.e. $v(0)\rho(0,t)=0$, we get
\begin{equation}
\label{cpr}
C(t)=v(1)\rho(1,t).
\end{equation}
Now, the source term can be solved by applying Eqs. (\ref{rhoh}) and (\ref{nhom}) in Eq. (\ref{cpr}) resulting
\begin{equation}
C(t)=\psi(t+\phi(1))+\int_{0}^{t}d\tau\,C(\tau)\psi(t-\tau+\phi(1)).
\label{stlong}
\end{equation}
Applying the Laplace transform and the convolution theorem in Eq. (\ref{stlong}) one obtains
\begin{equation}
\tilde{C}(s)=\frac{\tilde{\psi}(s+\phi(1))}{1-\tilde{\psi}(s+\phi(1))}.
\label{stlap}
\end{equation}
Finally, $\tilde{\rho}(x,s)$ is obtained by using convolution theorem in Eq. (\ref{nhom}), leading to the solution (\ref{solve}).

\subsection*{\bf A2. Conservation of $\int_{0}^{1}\rho(x,t)dx$}

We can check the conservation of measure $\int_{0}^{1}\rho(x,t)dx$ from general solution (\ref{solve}). First, Eq. (\ref{psi}) reads
\begin{equation}
\label{prco}
\tilde{\psi}(s+\phi(x))=x-s\mathcal{L}_{s}[\phi^{-1}(t+\phi(x))].
\end{equation}
Now, from Eqs. (\ref{phi}), (\ref{psi}), and (\ref{prco}) we have
\begin{eqnarray}
\label{cons}
\int_{0}^{1}\frac{1}{v(x)}\tilde{\psi}(s+\phi(x))dx&=&-\int_{0}^{1}\tilde{\psi}(s+\phi(x))\phi'(x)dx\nonumber\\
&=&\int_{0}^{\infty}e^{-st}\left[-\int_{0}^{1}\psi(t+\phi(x))\phi'(x)dx\right]dt\nonumber\\
&=&\int_{0}^{\infty}e^{-st}\left\{\int_{0}^{1}\frac{\partial}{\partial t}[\phi^{-1}(t+\phi(x))]\phi'(x)dx\right\}dt\nonumber\\
&=&\int_{0}^{\infty}e^{-st}\left[\int_{0}^{1}\frac{\partial}{\partial x}\phi^{-1}(t+\phi(x))dx\right]dt\nonumber\\
&=&\mathcal{L}_{s}[\phi^{-1}(t)]=\frac{1}{s}\left[1-\tilde{\psi}(s)\right],
\end{eqnarray}
noting that $\phi(1)=0$ and $\phi^{-1}(t+\phi(0))=0$ since $\phi(0)=\infty$. Finally, Eqs. (\ref{solve}) and (\ref{cons}) give us
\begin{equation}
\label{fk}
s\int_{0}^{1}\tilde{\rho}(x,s)dx-1=\mathcal{L}_{s}\left[\frac{d}{dt}\int_{0}^{1}\rho(x,t)dx\right]=0,
\end{equation}
recalling that $\rho(x,0)=1$.

\subsection*{\bf A3. Asymptotic solution}

The scaling $\rho(x,t)\sim\zeta'(t)\rho(x)$ is just the $s=0$ ($t\rightarrow\infty$) lowest order expansion of $\tilde{\rho}(x,s)$ near $x=0$: from Eqs. (\ref{solve}) and (\ref{sepwe}) one has 
\begin{equation}
\label{scla}
\tilde{\rho}(x,s)\sim s\tilde{\zeta}(s)\rho(x)\lim_{x\rightarrow0}\frac{\tilde{\psi}[\phi(x)]}{\rho(x)v(x)},\qquad s\rightarrow0,
\end{equation}
while Eq. (\ref{prco}) reads
\begin{equation}
\label{psta}
\tilde{\psi}[\phi(x)]=x-\lim_{s\rightarrow0}s\mathcal{L}_{s}[\phi^{-1}(t+\phi(x))].
\end{equation}
From Eqs. (\ref{wcc}) and (\ref{zetaf}) we have $\phi^{-1}(t)\sim t^{-\gamma}/q(t)$ and, by making use of Karamata's Abelian and Tauberian theorems, the dependence of Eq. (\ref{psta}) on $s$ is as follows
\begin{equation}
\label{ptaub}
s\mathcal{L}_{s}[\phi^{-1}(t+\phi(x))]\sim O(s^{\gamma}/l(1/s)),\qquad s\rightarrow0,
\end{equation}
and thus $\tilde{\psi}[\phi(x)]=x$. Recalling that $\rho(x)v(x)\sim x$ from Eq. (\ref{vxrho}), we finally get the scaling relation previously proposed.

\end{document}